# Design and Implementation of a Micron-Sized Electron Column Fabricated by Focused Ion Beam Milling

Flavio Wicki, Jean-Nicolas Longchamp, Conrad Escher, Hans-Werner Fink

Physics Department, University of Zurich

Winterthurerstrasse 190, 8057 Zurich, Switzerland

Corresponding author: Flavio Wicki, flavio.wicki@physik.uzh.ch

**Highlights**

- Electron Optics
- Scaling laws
- Low-energy electrons
- Coherent electron beams
- Micron-sized electron column

**Abstract**

We have designed, fabricated and tested a micron-sized electron column with an overall length of about 700 microns comprising two electron lenses; a micro-lens with a minimal bore of 1 micron followed by a second lens with a bore of up to 50 microns in diameter to shape a coherent low-energy electron wave front. The design criteria follow the notion of scaling down source size, lens-dimensions and kinetic electron energy for minimizing spherical aberrations to ensure a parallel coherent electron wave front. All lens apertures have been milled employing a focused ion beam and could thus be precisely aligned within a tolerance of about 300 nm from the optical axis. Experimentally, the final column shapes a quasi-planar wave front with a minimal full divergence angle of 4 mrad and electron energies as low as 100 eV.

**Keywords**

Electron optics, low-energy electrons, imaging, coherent diffraction, electron lens

## 1. Introduction

In past decades, electron microscopy has made significant progress, in particular by the realisation and implementation of aberration correcting elements [1-2]. This has led to commercial transmission electron microscopes that routinely deliver atomic resolution in material science research. A current trend is to reduce beam energies from 100 keV, a typical value 10 years ago, down to energies in the several 10 keV regime [3-4]. The goal is to minimize knock-out damage and thus tolerate a reasonably high electron dose to achieve good signal to noise ratios and to even envision acquiring spectroscopic data on a single atom level.

Even lower energies in the 100 eV range are employed to operate microscopes with high surface sensitivity which has led to the impressive technology of LEEM pioneered by Telieps and Bauer some 30 years ago [5]. Recent versions of these surface sensitive tools are also equipped with aberration correctors to push the lateral resolution limit to or even below the nanometer range [6-7]. Related devices, also using a cathode lens close to the sample to decelerate the electron beam, are lately used in the SEM and STEM modes [8-9].

Comparably few efforts have been made in the development of electron microscopes operating in the 100 eV regime by which the electron energy is kept low throughout the entire electron column [10-13] instead of decelerating the beam just where the low energy is needed, close to the sample. Such efforts require scaled down electron lenses and enable technologies like coherent diffraction with low-energy electrons which has recently entered the 2 Angstrom resolution regime in imaging freestanding graphene [14].

Scaling down lens dimensions while maintaining atomic source size [15] and low emission voltages right at the field emission tip level leads to reduced lens aberrations. This concept is illustrated in Fig.1. For reaching a scaling factor of 1000 or more, lenses need to be machined with sub-micron precision.

If the geometrical dimensions of a given electrostatic electron lens are scaled down by a constant factor $k$ while the ratio of the electrode potentials to the beam energy remains unchanged, the run of the electron trajectories is scaled down by the same factor $k$. The geometrical aberrations given as deviations of the ray trajectories from the optical axis in the Gaussian image plane thus scale with $k$ as well.

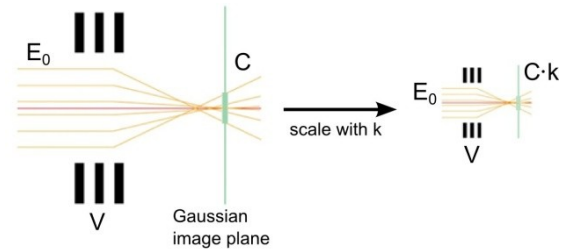

**Fig. 1.** Downsizing approach for low aberrations in electrostatic electron lenses. When scaling down the size of a lens by a factor $k$, the geometrical aberrations are reduced by that same factor, provided that the ratio of the electron energy $E_0$ to the electrode potential $V$ is preserved. $C$ denotes the transversal spherical aberration of the lens.

## 2. Concept

In coherent diffraction imaging (CDI), micro-fabricated electron lenses have proven to be a suitable tool to shape a spherical wave front emitted from a field emission tip into a nearly parallel one [16-17]. A micron-sized two electrode electron lens (micro-lens) has already been successfully employed in CDI experiments [14]. To get more control over the beam properties, we took the effort to incorporate a second electron optical element placed behind such a micro-lens. Sticking to the downsizing approach, an electrostatic three electrode lens (einzel-lens) with small dimensions was chosen as the second element to keep the aberrations low.

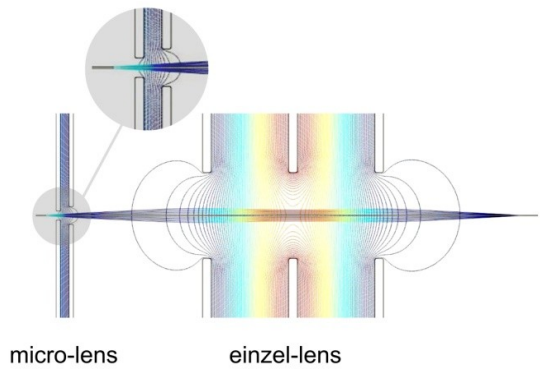

micro-lens     einzel-lens

**Fig. 2.** Illustration of the working principle of the micro-column. Potential differences applied between the two micro-lens electrodes and between the middle and outer einzel-lens electrodes create an electrostatic field distribution which has a focusing effect on the divergent electron beam emitted by the field emission source. The inset shows an enlarged view of the micro-lens apertures.

The setup of this five electrode electron optical column (micro-column) is illustrated in Fig. 2. It consists of a micro-lens with aperture diameters in the range of 1 to 5 µm, followed by an einzel-lens with a maximum aperture diameter of 50 µm. The micro-lens acts as both, beam limiting aperture and extractor for field electron emission. If a potential difference between the two electrodes is applied, an electrostatic field distribution around the apertures is formed, which has a focusing effect on a penetrating electron beam.

Pre-shaped like this, the electron beam subsequently enters the einzel-lens. The outer electrodes of the latter are usually kept at ground potential and either a negative (retarding mode) or a positive voltage (accelerating mode) is applied to the middle electrode. In either mode, the electrons experience a net acceleration towards the optical axis and provided that the voltage relative to the electron energy is high enough, the electron beam converges.

In order to control the beam energy throughout the column, the electron emitter as well as the micro-lens can be biased with respect to the einzel-lens. This is of particular relevance for diffraction experiments in view of tuning the wavelength of the electrons.

## 3. Micro-column fabrication procedure

The building blocks for the micro-column fabrication are commercially available 100 nm thick silicon nitride membranes covering a 250x250 µm$^2$ window in a 100 µm thick silicon substrate. According to the number of electrodes, the micro-lens and the einzel-lens are pre-assembled to stacks of two and three building blocks respectively, using vacuum compatible epoxy. The fabrication procedure is summarized in Fig. 3.

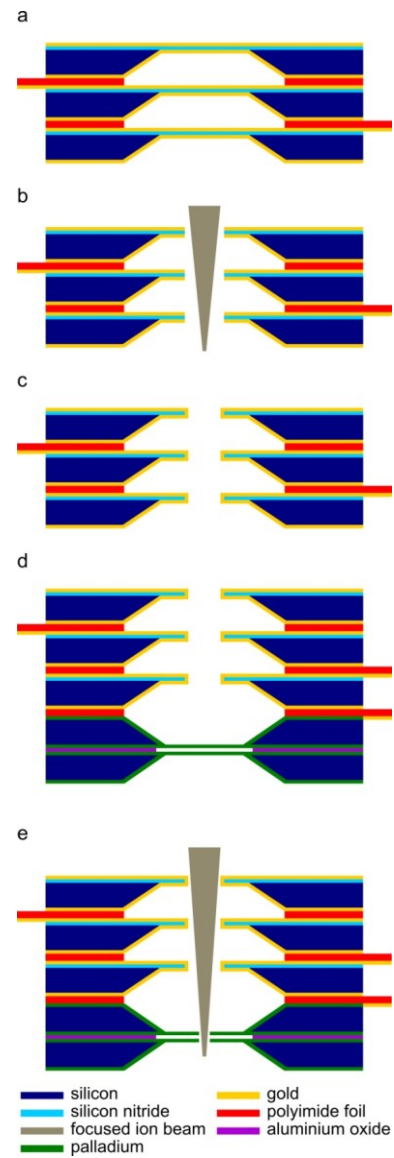

■ silicon    ■ gold
■ silicon nitride    ■ polyimide foil
■ focused ion beam    ■ aluminium oxide
■ palladium

**Fig. 3.** Illustration of the micro-column fabrication procedure. (a) Three silicon chips are stacked for the einzel-lens fabrication. (b) The apertures of the einzel-lens are milled using a focused gallium ion beam. (c) The apertures are sputter coated with gold to cover the silicon nitride revealed during milling. (d) The micro-lens stack is bonded to the einzel-lens. (e) Finally the micro-lens apertures are milled via the einzel-lens apertures.

### 3.1. Einzel-lens

For the electrode fabrication of the einzel-lens, all three building blocks are first coated with gold on both sides. In order to electrically insulate the individual electrodes within the stack, they are seperated by a 75 µm thick polyimide foil leading to a total membrane to membrane spacing of about 200 µm. The polyimide spacers are gold coated on one side and hence allow for electrically contacting the electrodes. Once such stack of spacers and membranes is assembled, all three apertures are ion milled through the membranes along the stack axis in a single run (see Fig. 3(a) and (b)). After milling, the silicon nitride rims of the apertures need sputter coating in order to prevent charging during operation (see Fig. 3(c)). SEM and scanning ion microscopy images of the einzel-lens are shown in Fig. 4.

### 3.2. Micro-lens

For the pre-assembly of the micro-lens electrodes, two building blocks are first coated with a 200 nm thick palladium layer on the upper surface only. Thereafter, the supporting silicon nitride windows are removed by reactive ion etching leaving freestanding palladium membranes behind. For subsequently contacting the electrodes, the building blocks are now additionally palladium coated on the bottom surface. Finally, a 1 µm thick aluminium oxide frame is electron beam evaporated around the palladium membrane on the upper side for electrical insulation. Prepared like this, the two building blocks are assembled with the aluminium oxide layers facing each other. This leads to a total membrane to membrane spacing of approximately 5 µm. SEM and scanning ion microscopy images of micro-lens apertures are shown in Fig. 5.

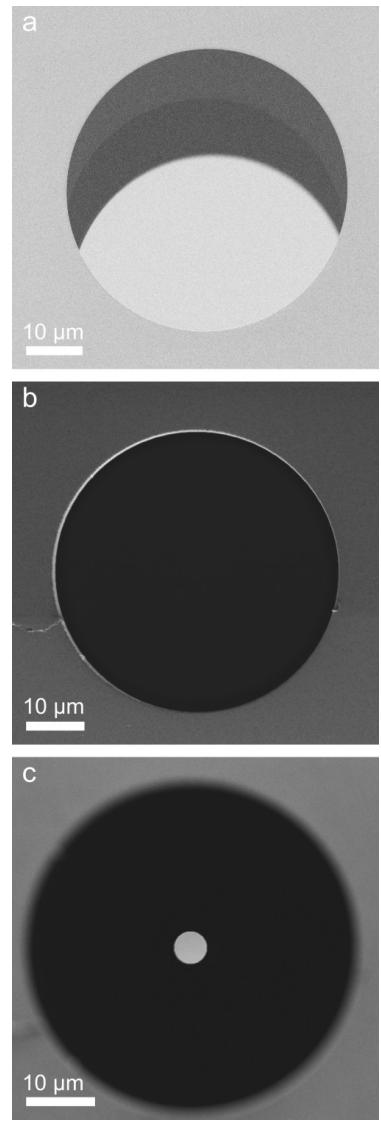

**Fig. 4.** Scanning electron and ion microscopy images of the micro-column elements. (a) SEM image recorded with the sample tilted to display the three einzel-lens apertures. (b) Scanning ion microscopy image of the einzel-lens exit aperture, illustrating its high circularity. (c) Scanning ion microscopy image of a micro-lens aperture with 5 µm in diameter viewed through the einzel-lens revealing accurate alignment of the two lenses. The focus is adjusted to the micro-lens.

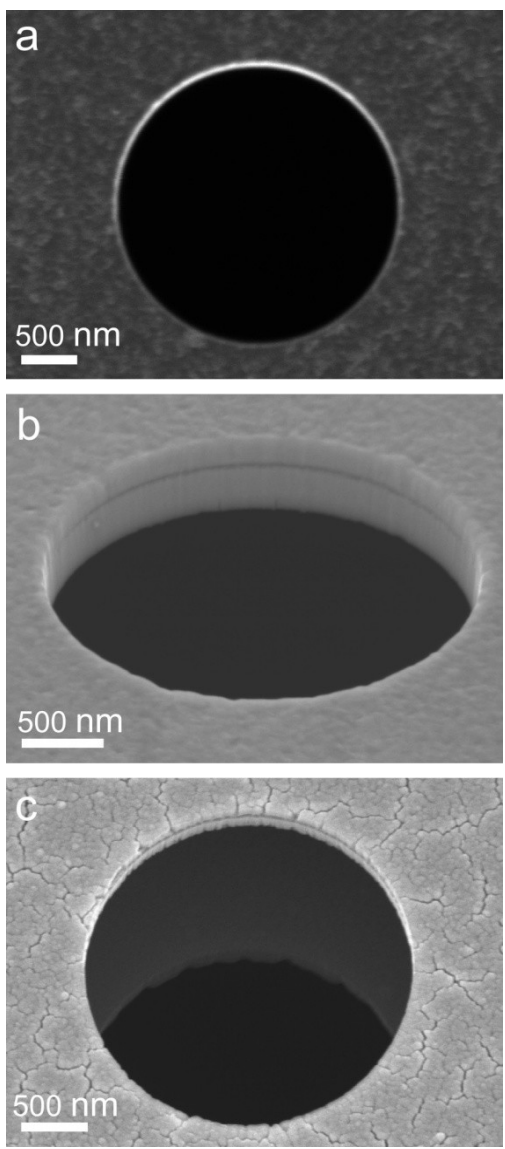

**Fig. 5.** Scanning electron and ion microscopy images of the micro-lens. (a) Scanning ion microscopy image of a 2.5 μm aperture in a freestanding palladium membrane, recorded right after milling. (b) SEM image of the aperture recorded under an angle of 54°. (c) SEM image of a micro-lens recorded under an angle of 23° presenting both apertures.

### 3.3. Micro-column

In order to build a micro-column, the micro-lens pre-assembly is attached onto the einzel-lens (see Fig. 3(d)). Again, the two lenses are insulated by a polyimide foil, gold coated on one side to provide electrical contact to the exit electrode of the micro-lens. The electrode spacing between exit of the micro-lens and inlet of the einzel-lens adds up to about 300 μm. The apertures in the palladium membranes of the micro-lens are finally milled from the exit of the column via the apertures of the einzel-lens (see Fig. 3(e)). In this way an alignment of both lenses within less than 300 nm can be attained (see Fig. 4(c)).

### 4. Experimental characterization

Prior to experimental tests under UHV conditions, the micro-column is mounted on a sample holder. In doing so, the electrodes are contacted using silver paint (see Fig. 6). An electrochemically etched W(111) field emission tip is used as source of a divergent coherent low-energy electron beam. The field emitter is mounted on a 3-axis piezo-positioner and can thus be precisely positioned in front of the micro-lens.

In order to monitor the performance of the micro-column, a detector unit featuring a microchannel plate followed by a phosphor screen is placed at a distance of 70 mm beyond the micro-colum. The signal on the screen is captured by a CCD camera with 6000x8000 pixels and 16 bit dynamic range.

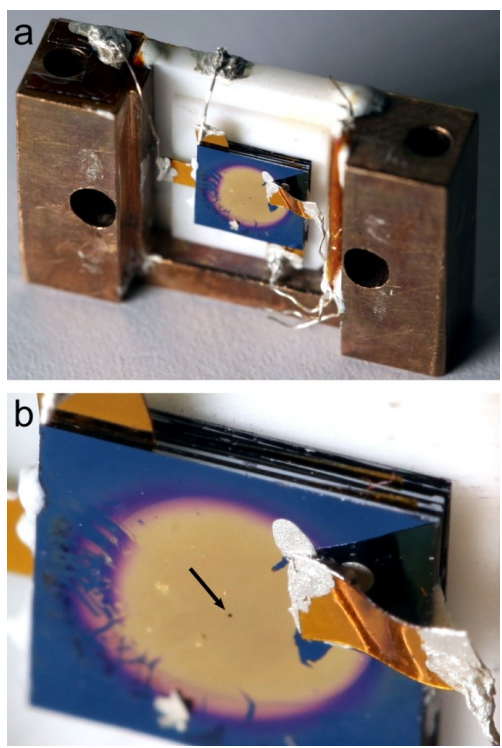

**Fig. 6.** Photographs of the mounted micro-column. (a) The micro-column is mounted on a sample holder for testing in an UHV system. The overall sample holder width amounts to 19 mm. The view is towards the exit side of the column. (b) Close up view of the column. The arrow points to the last 50 μm diameter einzel-lens aperture.

When all the micro-column electrodes are kept at ground potential, a projection image of the second micro-lens electrode aperture is recorded on the detector screen (see Fig. 7(a)). Since this is the beam limiting aperture of the system, the projection image implies the maximum acceptance angle of the micro-column. Due to the large distance of the detector from the micro-column compared to the maximum beam diameter inside the column, a minimal spot is recorded when the beam leaves the micro-column almost parallel. We estimate the divergence angle of the nearly parallel beam as the ratio of the spot size of the beam in the detector plane and the distance to the micro-column. This is also how the scale bar in Fig. 7(a)-(d) is calibrated, 30 mrad correspond to 2.1 mm on the detector. The divergence angle of the nearly parallel beam and the beam profile characterize the electron optical performance of the column.

Fig. 7 illustrates the performance of a micro-column consisting of a micro-lens with 2.5 μm apertures and an einzel-lens with 50 μm apertures. During the test series the field emitter is kept aligned at a certain distance in front of the micro-lens. The extraction voltage is set to -100 V.

The size of the projection image of the beam limiting aperture shown in Fig. 7(a) provides the acceptance angle of the micro-column. Reading out the FWHM of the linescan depicted in Fig. 8(a) results in a full divergence angle of 54 mrad, which corresponds to a distance between tip and beam limiting aperture of 46 μm.

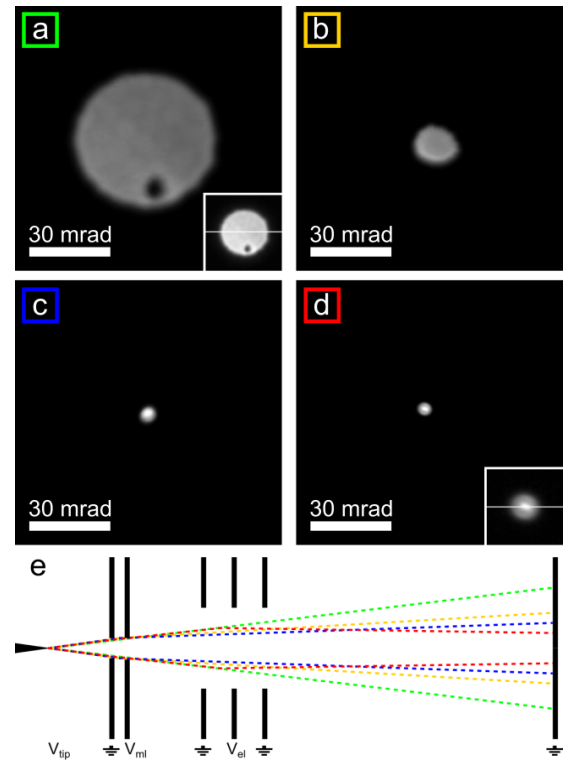

**Fig. 7.** Performance of a micro-column consisting of a micro-lens with 2.5 μm apertures and an einzel-lens with 50 μm apertures. (a)-(d) Beam spots recorded for different voltages $V_{ml}$ on the second micro-lens aperture and $V_{el}$ on the middle einzel-lens aperture. All the other lens electrodes are at ground potential while the tip potential is kept at $V_{tip}$ = -100 V. (a) Projection image of the second lens aperture ($V_{ml}$ = 0 V, $V_{el}$ = 0 V). The black spot visible inside the projection image is due to a defect on the detector. (b) The divergence angle of the electron beam is reduced by the micro-lens ($V_{ml}$ = -40 V, $V_{el}$ = 0 V). (c) Minimal beam spot in case solely the micro-lens is used to shape the beam ($V_{ml}$ = -53 V, $V_{el}$ = 0 V) (d) Minimal beam spot experimentally attained when micro- and einzel-lens are operated in combination ($V_{ml}$ = -40 V, $V_{el}$ = 160 V). (e) Sketch of the micro-column and the beam envelope for the different conditions discussed above: green corresponds to the condition shown in (a), orange (b), blue (c) and red (d).

The divergence angle of the beam is reduced when a potential of -40 V is applied to the second micro-lens electrode (see Fig. 7(b)), and as long as the micro-lens alone is in operation, a minimal spot is formed on the detector for a potential of -53 V, as shown in Fig. 7(c). The spot size on the detector can be further reduced, when micro-lens and einzel-lens are both in operation. The divergence angle of the beam is first reduced with the micro-lens, and the einzel-lens subsequently shapes the beam in the accelerating mode with a potential of +160 V on the middle electrode to the spot shown in Fig. 7(d). A linescan through the intensity distribution of the spot and a 3-dimensional representation are shown in Fig. 8(b) and (c). The FWHM of the linescan corresponds to a full divergence angle of 4 mrad.

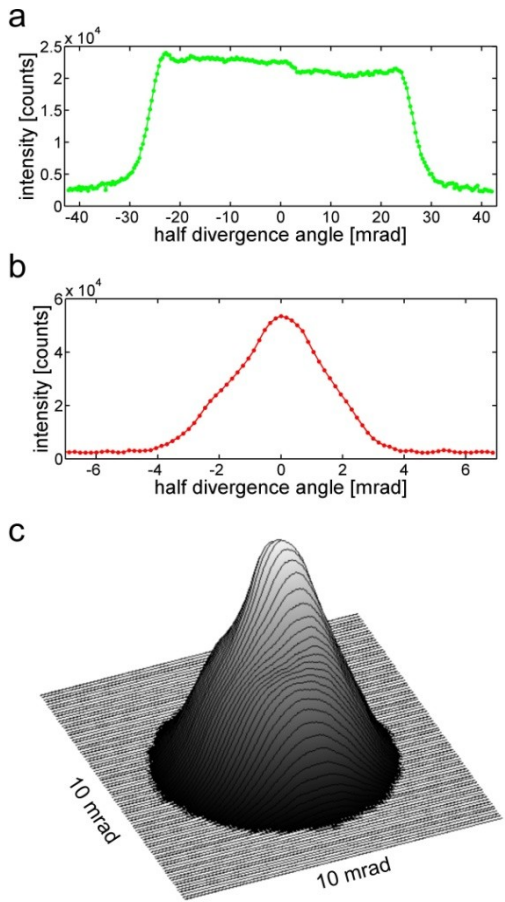

**Fig. 8.** (a) Intensity profile along the line shown in the inset in Fig. 7(a). (b) Intensity profile along the line shown in the inset in Fig. 7(d). (c) 3-dimensional representation of the beam intensity profile from the spot in Fig. 7(d).

## 5. Conclusions

By sheer downsizing of the lens dimensions, aberations intrinsic to conventional electrostatic electron lenses can be minimized to that extent where lens characteristics such as electrode alignment and rotational symmetry remain sufficiently precise. Using a gallium ion beam to successively mill apertures through stacks of membrane electrodes allows assembling a dual lens electron optical column featuring micron-sized apertures of high circularity on a well defined optical axis.

Adjusted accordingly, the electron column provides a nearly parallel beam with a residual minimal half divergence angle of 2 mrad. Most notably, the dual lens system allows accelerating and decelerating the electrons within the column. Consequently, for a given focal length electron energies can be selected from a range between a few eV up to more than 1 keV. This aspect again is very beneficial for CDI applications: a coherent parallel electron beam of tunable energy allows adjusting the spatial resolution contained in the diffraction record without reconditioning the acceptance angle of the detector unit.

## Acknowledgment

We would like to thank the Swiss National Science Foundation for financial support (grant numbers PZ00P2_148084 and 200021_150049).